\documentclass{article}
\usepackage{sita2005}

\title{
The 
exponential family of Markov chains \\ 
and its information geometry
}

\author{Hiroshi Nagaoka
  \thanks{
 Graduate School of Information Systems, 
The University of Electro-Communications.
1-5-1, Chouhugaoka, Chouhu-shi, Tokyo, 182-8585, Japan.
(e-mail: nagaoka@is.uec.ac.jp)
  }
}

\abstract{
We introduce a new definiton of exponential family of Markov chains, 
and show that 
many characteristic properties of the usual exponential family of 
probability distributions are properly extended to Markov chains.
The method of information geometry is effectively applied 
to our framework, which enables us to characterize the divergence 
rate of Markov chain from a differential geometric viewpoint. 
}


\usepackage{macro}
\usepackage{amsmath}
\usepackage[psamsfonts]{amssymb}

\begin{document}
\maketitle

\section{Introduction}
A $d$-dimensional parametric family 
\[ \condset{p_\theta}{\theta=
(\theta^1, \ldots , \theta^d)\in\Theta}, \quad \Theta\subset\bR^d
\]
of probability  density functions on 
a measure space $(\cX, {\cal B}, \nu) $ is called an exponential family with a 
natural parameter $\theta=(\theta^i)$ 
(e.g.\ \cite{BN:book}) 
when there exist $d+1$ $\bR$-valued functions 
$C, F_1, \ldots , F_d$ on $\cX$ and an $\bR$-valued 
function $\psi$ on $\Theta$ such that 
\begin{equation}
\label{eq:exp_family_p}
\log p_\theta (x) = 
C(x) + \sum_{i=1}^d \theta^i F_i(x) -\psi (\theta).
\end{equation}
The notion of exponential family is very important 
in various fields such as the theory of statistical inference (parameter estimation, 
hypothesis testing, etc.), large deviations, information theory, etc, 
and a dualistic viewpoint of the information geometry (\cite{Ama:book, 
AmaNag:book, NagAma82}) works effectively to understand 
the structure of exponential families in a unified and elegant manner. 
The present paper is aimed at demonstrating that many characteristic properties 
of exponential families (\ref{eq:exp_family_p}) are preserved by 
families of Markov kernel densities of the form
\begin{align}
\log w_\theta (y|x) = &
C(x,y) + \sum_{i=1}^d \theta^i F_i(x,y) 
\nonumber \\
& + K_\theta (y) - K_\theta (x) -\psi (\theta).
\label{eq:exp_family_w}
\end{align}
In particular, we show that the family is dually flat 
just as is the case for (\ref{eq:exp_family_p}), which 
enables us to apply the general theory for dually flat 
spaces to investigation of the information-geometrical 
structure of the family.  In order to avoid 
being involeved with 
functional analytic arguments and to concentrate upon 
geometric and algebraic aspects, we mostly confine ourselves 
to the case where $\cX$ is a finite set and $\nu$ is the counting 
measure, whereas the essence of the arguments can be extended to 
the general case by putting proper regurality conditions. 

Several attempts have been made so far 
to extend the definition of exponential familes to 
Markove processes or more general stochastic processes 
(e.g.\ \cite{Fei81, KucSor98, KucSor:book}).  Although 
they share the common criterion that an exponential 
family should have a finite dimensional (exact or asymptotic) sufficient 
statistic, the definitions given there are diverse, reflecting their 
respective backgrounds.  The authors of such papers as 
\cite{ItoAma88, Mer89, Ama01} had essentially the same 
concept on exponential families of Markov chains as ours when they said 
that the totality of strictly positive Markov chains of a fixed order 
is asymptotically regarded as an exponential family in an 
approximate sense.  We refine the concept so that the general 
definition of exponential families of Markov kernels is given 
without appealing to asymptotic settings, although the 
significance of the definition is made clear through some 
asymptotic arguments.  This non-asymptotic treatment 
enables us to develop the information geometry of 
Markov chains in a transparent and systematic way.  

This article is essentially based on the technical report \cite{NagISTR}. 
The proofs of the theorems are omitted for want of space.  

\section{Definition and an example} 

Let $\cX$ be a finite set with $|\cX|\geq 2$.  
We denote the totality of positive probability 
distributions on $\cX$ by $\cP=\cP(\cX)$. 
A Markov kernel (transition matrices) on $\cX$ 
is a map $w : \cX^2\rightarrow [0, 1]$ ($(x,y)\mapsto w(y|x)$) 
satisfying $\sum_{y\in\cX} w(y|x) =1$ for every $x\in\cX$. 
Suppose that we are given a subset $\cE$ of 
${\cX}^2 = \cX\times\cX$ for which the 
directed graph 
$(\cX, \cE)$ is strongly connected; i.e., for any $(x,y)\in{\cX}^2$ 
there exists a sequence $(x_1, x_2), (x_2, x_3), 
\ldots , (x_{n-1}, x_n)$ in $\cE$ such that 
$x_1=x, x_n=y, n\geq 2$.  Let the totality of 
irreducible 
Markov kernels $w$ on $\cX$ such that   
$\condset{(x,y)}{w(y|x)>0}=\cE$ be denoted by 
$\cW = \cW(\cX,\cE)$.  This includes the set of strictly positive 
Markov kernels $\cW(\cX,{\cX}^2)$ as a special case. 

From the irreducibility, each $w\in\cW$ has the unique 
stationary distribution in $\cP$, which is denoted by $p_w$. 
We introduce the notaion 
\[ p_w^{(n)}(x_1, \ldots, x_n)=
p_w(x_1) w(x_2|x_1)\cdots w(x_n|x_{n-1}) \]
($p_w^{(1)}(x)=p_w(x)$ and $p_w^{(2)}(x,y)=
p_w(x) w(y|x)$ in particular), 
which will be used in later sections.

A $d$-dimensional parametric
family 
\[ M=\condset{w_\theta}{\theta=(\theta^1, \ldots, \theta^d)\in\bRd} 
\subset \cW \]
is called an {\em exponential family} (or a {\em full 
exponential family} following the terminology of \cite{BN:book}) 
of Markov kernels on $(\cX,\cE)$ 
with 
a {\em natural (or canonical) parameter} $\theta$, when 
there exist functions $C, F^1, \ldots , F^d\, :\cX\rightarrow\bR$, 
$K : \cX\times\bRd\rightarrow\bR$ ($(x,\theta)\mapsto K_\theta(x)$) 
and $\psi:\bRd\rightarrow\bR$ such that equation (\ref{eq:exp_family_w}) 
holds for every $\theta\in\bRd$ and every $(x,y)\in\cE$.  

When a family of initial distributions, say $\condset{q_\theta}{\theta\in\bRd}$, 
is specified, the family of joint probability distributions for 
$x^n=(x_1, \ldots , x_n)$ satisfying $(x_t, x_{t+1})\in\cE$ ($\forall t$) 
is determined by
\begin{equation}
q^{(n)}_\theta (x^n) = q_\theta (x_1) w_\theta(x_2|x_1) 
\cdots w_\theta (x_n|x_{n-1}),
\end{equation}
for which we have from (\ref{eq:exp_family_w}) 
\begin{align}
\log q^{(n)}_\theta (x^n)  = &
\log q_\theta(x_1) + 
\sum_{t=1}^{n-1}
\Big\{
C(x_t, x_{t+1}) 
\nonumber \\
+ 
\sum_{i=1}^d &\theta^i F(x_t, x_{t+1}) - \psi(\theta)\Big\} 
+ K_\theta(x_n) - K_\theta(x_1) 
\nonumber \\
= &
\sum_{t=1}^{n-1} C(x_t, x_{t+1}) 
+
\sum_{i=1}^d \theta^i \sum_{t=1}^{n-1} F(x_t, x_{t+1}) 
\nonumber \\
& - (n-1) \psi (\theta) + O(1) 
\end{align}
This shows that the family $\condset{q^{(n)}_\theta}{\theta\in\bRd}$
is asymptotically an exponential family in the usual sense. 


\begin{example}
\label{example1}
\quad 
For any $w\in\cW = \cW (\cX, \cX^2)$, we have
\begin{align}
\nonumber
\log w(y|x) & =
\sum_{i=0}^{|\cX|-1}\sum_{j=1}^{|\cX|-1}
\log\frac{w(j|i) w(0|j)}{w(0|i) w(0|0)}
\, \delta_{i,j} (x,y) \\
+& \log w(0|x) - \log w(0|y) + \log w(0|0),
\end{align}
where $\delta_{i,j}(x,y) = 1$ if $(i,j) = (x,y)$ and $=0$ otherwise. 
We can easily verify this identity by respectively calculating the RHS for 
the two cases $y\neq 0$ and $y=0$.  This shows that $\cW$ is an 
exponential family with $\dim\cW=|\cX|\cdot(|\cW|-1)$. 
\end{example}

\section{Affine structures}

The gist of the definition will be clarified by a fundamental relation 
between the affine structure of 
functions on $\cE$ and the Markov kernels on $\cE$ as shown below. 
Let $\cF = \cF(\cX,\cE)$ be the totality of functions on $\cE$, 
which is a $|\cE|$-dimensional linear space. 
An element $f$ of $\cF$ is sometimes identified with its 
extension to a function on $\cX^2$ by letting $f(x,y)=0$ for 
any $(x,y)\in\cX^2\setminus\cE$.  For a $f\in\cF$ we define 
the functions $f(+,\cdot)$ and $f(\cdot,+)$ on $\cX$ by 
\[ f(+,x) = \sum_{y\in\cX} f(y,x) \quad\mbox{and}\quad
f(x,+) = \sum_{y\in\cX} f(x,y).
\]
A function $f\in\cF$ is said to be {\em shift-invariant} 
if $f(+,\cdot) = f(\cdot,+)$, and the totality of 
shift-invariant functions is denoted by 
$\cFS=\cFS (\cX,\cE)$. 
A function $\in\cF$ is said to be {\em anti-shift-invariant} 
if there exists a function $\kappa$ on $\cX$ such that 
$f(x,y) = \kappa(y) - \kappa(x)$ for any $(x,y)\in\cE$, 
and the totality of 
anit-shift-invariant functions is denoted by 
$\cFA=\cFA (\cX,\cE)$.  
The linear subspaces $\cFS$ and $\cFA$ of $\cF$ 
are the orthogonal 
complements of each other with respect to the inner 
product on $\cF$ defined by
$\langle f_1, f_2\rangle = \sum_{(x,y)\in\cE}
f_1(x,y) f_2(x,y)$, which follows from 
\begin{gather*}
\sum_{(x,y)\in\cE} f(x,y) \left(\kappa (y) -\kappa (x)\right) \\
= \sum_{x\in\cX} \left( f(+,x)-f(x,+)\right) \kappa (x). 
\end{gather*}
We thus have the direct sum decompositon $\cF= \cFS\oplus\cFA$.  In addition, 
from the assumption that $(\cX,\cE)$ is strongly 
connected, the necessary and sufficient condition for 
two functions $\kappa_1, \kappa_2$ to define the 
same element of $\cFA$ is that $\kappa_1-\kappa_2$ is 
constant on $\cX$.   This leads to $\dim\cFA = 
|\cX|-1$ and $\dim\cFS = \dim\cF-\dim\cFA= 
|\cE|-|\cX|+1$. 

Let 
\[
\cFpos =\cFpos (\cX,\cE)\defeq
\condset{f}{f:\cE\rightarrow\bR^+}
\subset \cF, 
\]
where $\bR^+= (0,\infty)$. 
Then we have the following theorem, which is 
a direct consequence of the Perron-Frobenius theorem
for irreducible nonnegative matrices. 

\begin{theorem}
{\em For any $f\in\cFpos$, there exist 
$w\in\cW$, $Z>0$ and $\gamma:\cX\rightarrow\bR^+$ such that 
\begin{equation}
\label{eq:normalization}
\forall (x,y)\in\cE, \;\; 
w(y|x)=\frac{1}{Z}\, \frac{\gamma(y)}{\gamma(x)}\, f(x,y).
\end{equation}
Here $w$ and $Z$ are unique, and $\gamma$ is unique 
up to a constant factor. 
}
\end{theorem}

Denoting the correspondence between $f$ and $w$ 
in (\ref{eq:normalization}) by $w=\Gamma (f)$, 
we define the mapping $\Gamma: \cFpos\rightarrow\cW$. 
Note that $\cW\subset\cFpos$ and $\Gamma(w)=w$ for 
any $w\in\cW$. 

Since $\exp(f(x,y))$ defines an element of $\cFpos$ for 
every $f$, we obtain a mapping  
\begin{align*} 
\Delta\defeq\Gamma\circ\exp \,:\, 
&\cF\rightarrow\cW \\
&f\mapsto \Gamma(\exp (f)).
\end{align*}
Noting that $w=\Delta(f)=\Gamma(\exp(f))$ is 
written as $w(y|x)=\exp(f(x,y))\gamma(y)/Z \gamma(x)$ 
from (\ref{eq:normalization}), we have 
\begin{equation}
\log w(y|x)=f(x,y)+\kappa(y)-\kappa(x) -\psi,
\end{equation}
where $\kappa (x) =\log \gamma(x)$ and $\psi=\log Z$. 
This means that $\log w\equiv f$ mod $\cFA\oplus\bR$, 
where a real constant is identified with the corresponding 
constant function in $\cF$.  Moreover, $\Delta$ 
gives a diffeomorphism from the quotient linear space 
$\cF /(\cFA\oplus\bR)$ to $\cW$.  Now we have the following theorem.

\begin{theorem}
{\em A subset $M$ of $\cW$ is an exponential family 
if and only if there exists an affine subspace $V$ of 
$\cF/(\cFA\oplus\bR)$ for which $M=\Delta (V)
\defeq\condset{\Delta(f)}{f\in V}$, and we have 
$\dim M=\dim V$.  Moreover, the correspondence 
between exponential familes and affine subspaces is 
one-to-one.
}
\end{theorem}

From this theorem, we have the following 
corollaries.

\begin{corollary}
{\em 
$\cW$ itself is an exponential family 
of dimension $|\cE|-|\cX|$.
}
\end{corollary}

Remember that in Example~\ref{example1} we have verified the 
above fact for the case of complete graph $\cE=\cX^2$ by a 
little complicated calculation.   

\begin{corollary}
{\em 
The 1-dimensional exponential family \\
$\condset{w_t}{t\in\bR}$ 
passing through 
given two kernels $w_0, w_1\in\cW$ is written 
in the form 
$w_t= \Gamma(w_1^t w_0^{1-t})$, 
or equivalently 
\begin{equation}
w_t(y|x)= \frac{1}{Z_t}\, \frac{\gamma_t(y)}{\gamma_t(x)}\, 
\{w_1(y|x)\}^t \{w_0(y|x)\}^{1-t} .
\end{equation}
}
\end{corollary}

A 1-dimensional exponential family is called an {\em e-geodesic}.

\begin{corollary}
{\em 
A subset $M$ of $\cW$ is an exponential family if and only if 
for any two points $w_0$ and $w_1$ in $M$ the e-geodesic 
$\condset{\Gamma(w_1^t w_0^{1-t})}{t\in\bR}$ lies in $M$.
}
\end{corollary}

\section{Fisher information}
For an arbitrary $d$-dimensional parametric family 
$M=\condset{w_\theta}{\theta=(\theta^i)\in\Theta}\subset 
\cW$, where $\Theta$ is an open subset of ${\bR}^d$, the 
Fisher information matrix $G_\theta=[g_{ij}(\theta)] :d\times d$ 
(with respect to the parameter $(\theta^i)$) 
is defined by
\begin{equation}
g_{ij}(\theta) \defeq 
\sum_{(x,y)\in\cE}p_\theta(x,y) \left\{\partial_i \log w_\theta(y|x)\right\} 
\left\{\partial_j \log w_\theta(y|x)\right\},
\end{equation}
where $p_\theta(x,y)=p_{w_\theta}^{(2)}(x,y)=p_{w_\theta}(x) w_\theta(y|x)$ 
and $\partial_i=\partial/\partial\theta^i$.  This definition is commonly 
used because of the following fact.  
Let $\condset{q_\theta}{\theta\in\Theta}$ be 
an arbitrary family of probability distributions on 
$\cX$ (possibly being independent of $\theta$) 
and consider the joint distributions  
\[
q^{(n)}_\theta (x^n) = q_\theta (x_1) w_\theta(x_2|x_1) 
\cdots w_\theta (x_n|x_{n-1}).
\]
Letting $G^{(n)}_\theta=[g^{(n)}_{ij}(\theta)]$ be 
the Fisher information matrix (in the usual sense) 
of the family $\condset{q^{(n)}_\theta}{\theta\in\Theta}$, 
we have
\begin{equation}
g_{ij}(\theta)=\lim_{n\rightarrow\infty}
\frac{1}{n} g^{(n)}_{ij}(\theta).
\end{equation}
From the information-geometric viewpoint, 
the Fisher information is regarded as a Riemannian 
metric $g$ through the relation $g(\partial_i, \partial_j)
=g_{ij}$ 
and is called the Fisher metric.

\begin{theorem}
{\em 
When $M=\condset{w_\theta}{\theta\in\bR^d}$ is 
an exponential family of the form (\ref{eq:exp_family_w}), 
the Fisher information matrix with respect to the 
natural parameter $(\theta^i)$ is given by
\begin{equation}
g_{ij}(\theta) = \partial_i \partial_j \psi (\theta) .
\end{equation}
}
\end{theorem}

\section{Expectation parameters}
For an exponential family 
$M=\condset{w_\theta}{\theta\in\bR^d}$ of 
the form  (\ref{eq:exp_family_w}), we define 
\begin{equation}
\eta_i(\theta)\defeq 
\sum_{(x,y)\in\cE} p_{w_\theta}^{(2)}(x,y) F_i(x,y) .
\end{equation}
Then $\eta=(\eta_1, \ldots , \eta_d)$ forms 
another coordinate system for $M$, which we call 
the {\em expectation parameter} corresponding to 
the representation (\ref{eq:exp_family_w}).

\begin{theorem}
{\em 
We have:
\begin{gather}
\eta_i = \partial_i\psi, \\
\partial_i\eta_j = g_{ij}, \\
\partial^i\theta^j= g^{ij}, \\
 [g^{ij}]=[g_{ij}]^{-1},
\end{gather}
where $\partial^i = \partial/\partial\eta_i$, and 
$[g^{ij}]$ denotes the Fisher information matrix 
with respect to the dual parameter $(\eta_i)$. 
Moreover, if we define $\varphi = \sum_i \theta^i\eta_i
- \psi$, we have
\begin{gather}
\theta^i = \partial^i\varphi, \\
g^{ij}=\partial^i\partial^j\varphi.
\end{gather}
}
\end{theorem}

\section{A dually flat structure}
For an arbitrary $d$-dimensional parametric family 
$M=\condset{w_\theta}{\theta=(\theta^i)\in\Theta}\subset 
\cW$, the {\em exponential connection} (or {\em e-connection} for short) 
$\nabla^{(\e)}$ and the {\em mixture connection} (or {\em m-connection} 
for short) 
$\nabla^{(\m)}$ are defined as follows.
\begin{align}
\Gamma^{(\e)}_{ij,k} &= 
g(\nabla^{(\e)}_{\partial_i}\partial_j, \partial_k) 
\nonumber \\
& = 
\sum_{(x,y)\in\cE} \partial_i \partial_j\log w_\theta (y|x) 
\partial_k p_\theta (x,y), 
\\
\Gamma^{(\m)}_{ij,k} &= 
g(\nabla^{(\m)}_{\partial_i}\partial_j, \partial_k) 
\nonumber \\
& = 
\sum_{(x,y)\in\cE} \partial_i \partial_j p_{\theta}(x,y) 
\partial_k \log w_\theta (y|x),
\end{align}
where $g$ is the Fisher metric. 
Then $\nabla^{(\e)}$ and $\nabla^{(\m)}$ are 
dual with respect to $g$ in the sense that $X g(Y, Z) = 
g(\nabla^{(\e)}_X Y, Z) + g(Y, \nabla^{(\m)}_X Y)$ holds for 
any vector fields $X, Y, Z$. 

\begin{theorem}
{\em 
For an exponential family 
$M=$ \\ $\condset{w_\theta}{\theta\in\bR^d}$ of 
the form  (\ref{eq:exp_family_w}), both 
$\nabla^{(\e)}$ and $\nabla^{(\m)}$ are flat, 
and $[\theta^i]$ and $[\eta_i]$ are 
affine coordinate systems of these connections, 
respectively. 
}
\end{theorem}

\begin{theorem}
{\em 
Suppose that $M$ is an exponential family 
and $N$ is a submanifold of $M$.  Then 
$N$ is an exponential family if and only if 
$N$ is auto-parallel with respect to the 
e-connection of $M$.
}
\end{theorem}

Let
\begin{align*}
\cP^{(2){\rm S}}  & =  \cP(\cE)\cap \cFS \\
= &
\condset{p^{(2)}\in\cP(\cE)}{\forall y, 
\sum_x p^{(2)}(x,y) = \sum_xp^{(2)}(y,x)}.
\end{align*}
Then $w\mapsto p^{(2)}_w$ gives a 
diffeomorphism from $\cW$ to $\cP^{(2){\rm S}}$.

\begin{theorem}
{\em 
The m-connection of $\cW$ is the natural flat connection 
induced from the convexity of $\cP^{(2){\rm S}}$.  
In particular, the m-geodesic (i.e., the auto-parallel 
curve with respect to the m-connection) connecting 
given two points $w_0$ and $w_1$ in $\cX$ is 
represented as
\begin{equation}
p^{(2)}_{w_t} = t \, p^{(2)}_{w_1} + (1-t) \, p^{(2)}_{w_0}.
\end{equation}
}
\end{theorem}

\section{Canonical Divergence}
Let $M$ be an exponential family of the form 
(\ref{eq:exp_family_w}).  Then the canonical 
divergence $D: M\times M\rightarrow\bR$ 
with respect to the dually flat structure
$(g, \nabla^{(\m)}, \nabla^{(\e)})$ is defined by
\begin{equation}
D(w_1\,|\,w_2) = 
\varphi (w_1) + \psi(w_2) - \sum_{i=1}^d 
\eta_i(w_1) \theta^i(w_2).
\end{equation}
The divergence $D$ is also characterized by the 
following property: let $\gamma$ be an m-geodsic
such that $\gamma(1)=w_1$ and $\gamma(0)=w_2$, 
and $\delta$ be an e-geodsic
such that $\delta(1)=w_3$ and $\delta(0)=w_2$, 
which intersect at $w_2$.  Then we have
\begin{equation}
D(w_1\,|\,w_2) + D(w_2\,|\,w_3) - D(w_1\,|\,w_3)
= g(\dot{\gamma}(0), \dot{\delta}(0)),
\end{equation}
where the RHS means the inner product between 
the tangent vectors of $\gamma$ and $\delta$ at 
the intersecting point $w_2$. 

\begin{theorem}
{\em 
$D$ is represented as
\begin{equation}
D(w_1\,|\,w_2) = 
\sum_{(x,y)\in\cE} p_{w_1}^{(2)}(x,y) \log \frac{w_1(y|x)}{w_2(y|x)}.
\end{equation}
}
\end{theorem}

This is nothing but the divergence rate of Markov chains.  
Actually, we have
\begin{equation}
D(w_1\,|\,w_2) = \lim_{n\rightarrow\infty}
\frac{1}{n} D(q_1^{(n)}\,|\,q_2^{(n)}),
\end{equation}
where $q_i^{(n)}$, $i=1, 2$ are defined as
\[ q_i^{(n)}(x_1, \ldots , x_n) = 
q_i(x_1) w(x_2|x_1)\cdots w(x_n|x_{n-1})
\]
by arbitrary distributions $q_i$ on $\cX$.

\section{Remaining subjects}

The following subjects can also be treated in the present 
framework or its obvious extension.

\begin{itemize}
\item Application to the large deviation theory.  
\item Some variant of Cram\'{e}r-Rao inequality, and an 
estimation-theoretic characterization of 
exponential families.
\item Extension to higher-order Markov chains,
and a hierarchy of exponential familes:
$\cP=\cW_0\subset \cW=\cW_1\subset \cW_2\subset \cdots $.
\item Extension to general measureble spaces, and autoregressive 
models as an example.
\end{itemize}

\end{document}